\documentclass[a4paper,twocolumn,11pt,accepted=2022-12-28]{quantumarticle}
\pdfoutput=1
\usepackage{graphicx}
\usepackage{color}
\usepackage{tabularx}
\usepackage{dcolumn}
\usepackage{longtable}
\usepackage{tensor}
\usepackage[utf8]{inputenc}
\usepackage[english]{babel}
\usepackage[T1]{fontenc}
\usepackage{hyperref}
\usepackage[numbers,sort&compress]{natbib}
\usepackage{placeins}
\usepackage{bm}
\usepackage{amsmath}
\usepackage{amsfonts}
\usepackage{amssymb}
\usepackage{textcomp, gensymb}
\providecommand{\U}[1]{\protect\rule{.1in}{.1in}}

\newcommand{\eq}[1]{Eq.~(\ref{#1})} %
\def\be{\begin{equation}} %
\def\ee{\end{equation}} %
\newcommand{\bea}{\begin{eqnarray}}
\newcommand{\eea}{\end{eqnarray}}
\newcommand{\HU}{\hat U}

\begin{document}
\title{Fluid fermionic fragments for optimizing quantum measurements of electronic Hamiltonians in the variational quantum eigensolver}
\author{Seonghoon Choi}
\affiliation{Department of Physical and Environmental Sciences, University of Toronto Scarborough, Toronto, Ontario M1C 1A4, Canada}
\affiliation{Chemical Physics Theory Group, Department of Chemistry, University of Toronto, Toronto, Ontario M5S 3H6, Canada}
\author{Ignacio Loaiza}
\affiliation{Department of Physical and Environmental Sciences, University of Toronto Scarborough, Toronto, Ontario M1C 1A4, Canada}
\affiliation{Chemical Physics Theory Group, Department of Chemistry, University of Toronto, Toronto, Ontario M5S 3H6, Canada}
\author{Artur F. Izmaylov}
\email{artur.izmaylov@utoronto.ca}
\affiliation{Department of Physical and Environmental Sciences, University of Toronto Scarborough, Toronto, Ontario M1C 1A4, Canada}
\affiliation{Chemical Physics Theory Group, Department of Chemistry, University of Toronto, Toronto, Ontario M5S 3H6, Canada}

\maketitle
\begin{abstract}
Measuring the expectation value of the molecular electronic Hamiltonian is one of the challenging parts of the variational quantum eigensolver. A widely used strategy is to express the Hamiltonian as a sum of measurable fragments using fermionic operator algebra.
Such fragments have an advantage of conserving molecular symmetries that can be used for error mitigation. The number of measurements required to obtain the Hamiltonian expectation value is proportional to a sum of fragment variances. Here, we introduce a new method for lowering the fragments' variances by exploiting flexibility in the fragments' form. Due to idempotency of the occupation number operators, some parts of two-electron fragments 
can be turned into one-electron fragments, which then can be partially collected in a purely one-electron fragment. This repartitioning does not affect the expectation value of the Hamiltonian but has non-vanishing contributions to the variance of each fragment. The proposed method finds the optimal repartitioning by employing variances estimated using a classically efficient proxy for the quantum wavefunction. Numerical tests on several molecules show that repartitioning of one-electron terms lowers the number of measurements by more than an order of magnitude. 
\end{abstract}
\graphicspath{{./Figures/}}

\section{Introduction}
\label{sec:intro}
The variational quantum eigensolver (VQE) \cite{Peruzzo_OBrien:2014,McClean_Aspuru-Guzik:2016,Rybinkin_Izmaylov:2020,Cerezo_Coles:2021,Anand_Aspuru-Guzik:2022} is a promising hybrid quantum-classical algorithm for finding the ground-state of the molecular electronic Hamiltonian,
\begin{equation}
\hat{H} = \sum_{pq}^{N} h_{pq} \hat{E}_{p}^{q} + \sum_{pqrs}^{N} g_{pqrs} \hat{E}_{p}^{q} \hat{E}_{r}^{s}, \label{eq:Ham_spin_orb} 
\end{equation}
presented here in so-called chemists' notation, i.e., in terms of one-electron excitation operators, $\hat{E}_{p}^{q} = \hat{a}_{p}^{\dagger} \hat{a}_{q}$; while $h_{pq}$ and $g_{pqrs}$ are related to one- and two-electronic integrals, $N$ is the number of spin-orbitals.

VQE circumvents the hardware limitations of today's noisy intermediate-scale quantum (NISQ) devices \cite{Preskill:2018} by exploiting both quantum and classical computers: A quantum computer prepares a parameterized trial state $|\psi_{\theta} \rangle$ and measures its energy $E_{\theta} = \langle \psi_{\theta} | \hat{H} | \psi_{\theta} \rangle$, while a classical computer suggests new $\theta$ parameters to minimize $E_{\theta}$ (for brevity, we omit $\theta$ from now). However, measuring the expectation value $E$ is not trivial because digital quantum computers can only measure polynomial functions of Pauli-$\hat{z}$ operators.

A class of widely used approaches based on fermionic operator algebra \cite{Berry_Babbush:2019, Motta_Chan:2021, Huggins_Babbush:2021, Yen_Izmaylov:2021, Cohn_Parrish:2021} measures $E$ by re-expressing the molecular electronic Hamiltonian in Eq.~(\ref{eq:Ham_spin_orb}) as 
\begin{align}
\hat{H} &= \hat{H}_{0} + \sum_{\alpha=1}^{N_{f}} \hat{H}_{\alpha} \nonumber \\ 
        &= \sum_{pq}^{N} h_{pq} \hat{E}_{p}^{q} + \sum_{\alpha=1}^{N_{f}} \sum_{pqrs} g_{pqrs}^{(\alpha)} \hat{E}_{p}^{q} \hat{E}_{r}^{s}  \nonumber \\
        &= \hat{U}_{0}^{\dagger} \left( \sum_{p}^{N} \lambda_{p} \hat{n}_{p} \right) \hat{U}_{0} \nonumber \\ & \hspace{57pt} + \sum_{\alpha=1}^{N_{f}} \hat{U}_{\alpha}^{\dagger} \left( \sum_{pq}^{N} \lambda_{pq}^{(\alpha)} \hat{n}_{p} \hat{n}_{q} \right) \hat{U}_{\alpha}, \label{eq:frag_ham}
\end{align}
where $\hat{n}_{p} = \hat{a}_{p}^{\dagger} \hat{a}_{p}$ are occupation number operators, and $\hat{U}_{\alpha} = \exp[\sum_{p > q}^{N} \theta_{pq}^{(\alpha)} (\hat{E}_{p}^{q} - \hat{E}_{q}^{p})]$ are orbital rotations. Since $\hat{n}_{p}$ is mapped to polynomial functions of Pauli-$\hat z$ under all standard 
qubit-fermion transformations \cite{Bravyi_Kitaev:2002, Seeley_Love:2012}, each $\hat{U}_{\alpha} \hat{H}_{\alpha} \hat{U}_{\alpha}^{\dagger}$ is measurable on a quantum computer. 
Therefore, $E$ can be obtained by measuring each $\hat{H}_{\alpha}$ term separately using
\begin{equation}
E = \sum_{\alpha=0}^{N_{f}} \langle \psi | \hat{H}_{\alpha} | \psi \rangle = \sum_{\alpha=0}^{N_{f}} \langle \hat{U}_{\alpha} \psi | \hat{U}_{\alpha} \hat{H}_{\alpha} \hat{U}_{\alpha}^{\dagger} | \hat{U}_{\alpha} \psi \rangle. \label{eq:measurable_frags}
\end{equation}
Equation~(\ref{eq:measurable_frags}) shows that to measure each $\hat{H}_{\alpha}$, one first has to apply $\hat{U}_{\alpha}$ on $|\psi\rangle$. Fortunately, implementing $\hat{U}_{\alpha}$ on a quantum computer is efficient as it only requires $\binom{N}{2}$ two-qubit rotations and a gate depth of $N$ \cite{Kivlichan_Babbush:2018}. 

To find the fragments, $\hat{H}_{\alpha}$, starting from the Hamiltonian in the second-quantized form, one can employ either the low-rank (LR) decomposition \cite{Berry_Babbush:2019, Motta_Chan:2021, Huggins_Babbush:2021} or the full-rank (FR) optimization \cite{Yen_Izmaylov:2021, Cohn_Parrish:2021}. Both methods are only concerned with the two-electron fragments because $\hat{U}_{0}$ for the one-electron fragment ($\hat{H}_{0}$) can be found easily as it simply corresponds to a unitary matrix that diagonalizes the one-electron tensor, $h_{pq}$ \cite{Kivlichan_Babbush:2018}. The difference between LR and FR fragments is the rank of the resulting $\lambda_{pq}^{(\alpha)}$ [in Eq.~(\ref{eq:frag_ham})]. In the LR decomposition, $\lambda_{pq}^{(\alpha)}$ is an outer product of some vector $\eta_{p}^{(\alpha)}$ (i.e., $\lambda_{pq}^{(\alpha)} = \eta_{p}^{(\alpha)} \eta_{q}^{(\alpha)}$) and, therefore, has rank $1$. In contrast, 
in FR optimization, $\lambda_{pq}^{(\alpha)}$ can be a full-rank hermitian matrix. 
Increased flexibility of FR fragments was exploited in Ref.~\citenum{Yen_Izmaylov:2021} to lower the number of measurements required to obtain $E$ up to error $\epsilon$ when each $\hat{H}_{\alpha}$ is measured independently \cite{Crawford_Brierley:2021}:
\begin{equation}
M(\epsilon) = \frac{1}{\epsilon^{2}} \sum_{\alpha=0}^{N_{f}} \frac{\mathrm{Var}_{\psi}(\hat{H}_{\alpha})}{m_{\alpha}}, \label{eq:M_req}
\end{equation}
where $\mathrm{Var}_{\psi}(\hat{H}_{\alpha}) = \langle \psi| \hat{H}_{\alpha}^{2} |\psi \rangle - \langle \psi | \hat{H}_{\alpha} |\psi\rangle^{2}$ is the variance of $\hat{H}_{\alpha}$, and $m_{\alpha}$ is the fraction of the total measurements allocated to $\hat{H}_{\alpha}$. Developing techniques for measuring $E$ with a low $M(\epsilon)$ is especially important for VQE because a recent analysis \cite{Gonthier_Romero:2022} showed that the advantage of VQE over state-of-the-art classical algorithms is limited due to the large $M(\epsilon)$. 

Yet, even the best implementation of the FR optimization was shown to have higher $M(\epsilon)$'s than 
those in the measurement approaches developed in the qubit space \cite{Crawford_Brierley:2021, Jena_Mosca:2019, Huang_Preskill:2020, Hadfield_Mezzacapo:2022, Hilmich_Wille:2021, Huang_Preskill:2021, Wu_Yuan:2021, Hadfield:2021, Yen_Izmaylov:2020, Yen_Izmaylov:2022, Choi_Izmaylov:2022}. The qubit-space techniques start by applying one of the fermion-qubit mappings \cite{Bravyi_Kitaev:2002, Seeley_Love:2012} to the fermionic Hamiltonian to produce the qubit Hamiltonian,
\begin{equation}
\hat{H}_{q} = \sum_{j=1} c_{j} \hat{P}_{j}, \label{eq:H_q}
\end{equation}
where each $\hat{P}_{j}$ is an $N$-qubit Pauli product (i.e., a tensor product of Pauli operators for individual qubits). Thus obtained $\hat{H}_{q}$ is subsequently partitioned into a linear combination of $N_{f}$ independently measured fragments, $\hat{H}_{\beta}$ (i.e., $\hat{H}_{q} = \sum_{\beta=1}^{N_{f}} \hat{H}_{\beta}$), where
\begin{equation}
 \hat{H}_{\beta} = \hat{V}_{\beta}^{\dagger}\left( \sum_{i}a_{i}^{(\beta)} \hat{z}_{i} + \sum_{ij}a_{ij}^{(\beta)} \hat{z}_{i}\hat{z}_{j} + \dots \right) \hat{V}_{\beta}.
\end{equation}
Every $\hat{H}_{\beta}$ contains only mutually commutative Pauli products and thus allows one to efficiently implement 
the corresponding $\hat{V}_{\beta}$ using only one- and two-qubit Clifford gates \cite{Aaronson_Gottesman:2004,Crawford_Brierley:2021, Yen_Izmaylov:2020, Bansingh_Izmaylov:2022}. 

The qubit-space methods with the lowest $M(\epsilon)$ take advantage of the flexibility in the fragments  offered by the realization that some $\hat{P}_{j}$ can belong to multiple $\hat{H}_{\beta}$. The coefficients of $\hat{P}_{j}$ in different $\hat{H}_{\beta}$, $c_{j}^{(\beta)}$, can be varied without changing the total expectation value of $\hat{H}_{q}$ as long as $c_{j}^{(\beta)}$ sum to $c_{j}$ in the qubit Hamiltonian \cite{Yen_Izmaylov:2022}. In addition, even $\hat{P}_{j}$ not present in $\hat{H}_{q}$ can be introduced into multiple $\hat{H}_{\beta}$ provided that corresponding $c_{j}^{(\beta)}$ sum to zero \cite{Choi_Izmaylov:2022}. A significant reduction in $M(\epsilon)$ was achieved by optimizing $c_{j}^{(\beta)}$ using approximate variances obtained by employing a classically efficient wavefunction, $|\phi\rangle$, to estimate $\mathrm{Var}_{\phi}(\hat{H}_{\beta})$. The idea of increasing the number of $\hat{P}_{j}$ measured simultaneously in a single fragment has been successfully employed also in the recently developed classical-shadow-based techniques \cite{Huang_Preskill:2020, Hadfield_Mezzacapo:2022, Huang_Preskill:2021, Hilmich_Wille:2021, Wu_Yuan:2021, Hadfield:2021} to yield lower $M(\epsilon)$ values. An alternative class of promising qubit-space approaches with $M(\epsilon)$'s competitive with those in some of the Hamiltonian partitioning schemes lowers $M(\epsilon)$ by optimizing positive operator-valued measures (POVMs) \cite{Garcia-Perez_Maniscalco:2021, Fischer_Tavernelli:2022, Glos_Garcia-Perez:2022}.

In this work, we present an extension to fermionic fragment techniques that further increases their 
flexibility in reducing the number of measurements. The new approach generalizes the technique of 
repartitioning of some fragments used in the qubit space. It extends the repartitioning idea from 
commuting to non-commuting operators. 
Another motivation for developing fermionic measurement schemes is their advantage compared to the  
qubit-space counterparts in conserving molecular symmetries (e.g., 
electronic number and spin operators). These symmetries can be used for error mitigation 
techniques, which are essential for advancing quantum computing schemes on 
near-term devices \cite{Ryabinkin_cVQE:2019,Google_HF:2020}. 

\section{Theory}
\label{sec:theory}

\subsection{Fluid fermionic fragments}

Here, we present a new approach that exploits two properties of fermionic operators 
to minimize the number of measurements in \eq{eq:M_req}. First, any linear combination of one-electron hermitian operators 
can be brought to the factorized form
\begin{equation}
\sum_\alpha c_\alpha \HU_\alpha^\dagger \left( \sum_p \epsilon_p^{(\alpha)} \hat n_p \right) \HU_\alpha = 
\HU^\dagger \left( \sum_p \epsilon_p \hat n_p \right) \HU,
\end{equation} 
where $c_\alpha,\epsilon_p^{(\alpha)},$ and $\epsilon_p$ are some real coefficients, 
and $\HU_\alpha, \HU$ are orbital rotations. Second, the occupation number operators are idempotent, i.e.,
$\hat{n}_{p}^{2} = \hat{n}_{p}$. 
 
Using the $\hat{n}_{p}$ idempotency, each $\hat{H}_{\alpha}$ in \eq{eq:frag_ham} with $\alpha > 0$ can be re-written as a sum of one- and two-electron parts:
\begin{equation}
\hat{H}_{\alpha} = \hat{U}_{\alpha}^{\dagger} \left( \sum_{p}^{N} \lambda_{pp}^{(\alpha)} \hat{n}_{p} + \sum_{p\neq q} \lambda_{pq}^{(\alpha)} \hat{n}_{p} \hat{n}_{q} \right) \hat{U}_{\alpha}.\label{eq:H_alpha_oe_te}
\end{equation}
This expression reveals the freedom that one can extract any amount of the one-electron part from every $\hat{H}_{\alpha}$ and add it to $\hat{H}_{0}$, thereby repartitioning the one- and two-electron Hamiltonians. Thus, the repartitioned fragments, which we will refer to as fluid fermionic fragments (F$^3$), are
\begin{align}
\hat{H}_{0}^{\prime} &= \hat{H}_{0} + \sum_{\alpha=1}^{N_{f}} \hat{U}_{\alpha}^{\dagger} \left( \sum_{p}^{N} c_{p}^{(\alpha)} \hat{n}_{p} \right) \hat{U}_{\alpha}, \label{eq:mod_oe} \\
\hat{H}_{\alpha}^{\prime} &= \hat{H}_{\alpha} - \hat{U}_{\alpha}^{\dagger} \left( \sum_{p}^{N} c_{p}^{(\alpha)} \hat{n}_{p} \right) \hat{U}_{\alpha} \nonumber \\
  &\hspace{-15pt}= \hat{U}_{\alpha}^{\dagger} \left[ \sum_{p}^{N} (\lambda_{pp}^{(\alpha)}-c_{p}^{(\alpha)}) \hat{n}_{p} + \sum_{p\neq q} \lambda_{pq}^{(\alpha)} \hat{n}_{p} \hat{n}_{q} \right] \hat{U}_{\alpha} \label{eq:mod_te}.
\end{align}
Even after the modification, each $\hat{H}_{\alpha}$ for $\alpha>0$ remains measurable because $\hat{U}_{\alpha} \hat{H}_{\alpha} \hat{U}_{\alpha}^{\dagger}$ still maps onto a polynomial function of Pauli-$\hat{z}$ after qubit-fermion transformations. For $\hat{H}_{0}^{\prime}$, new $\hat{U}_{0}^{\prime}$ and $\lambda_{p}^{\prime}$ can easily be found by simply diagonalizing $h_{pq}^{\prime} = h_{pq} + \sum_{\alpha=1}^{N_{f}} \sum_{r}^{N} (U_{rp}^{(\alpha)})^{\ast} c_{r}^{(\alpha)} U_{rq}^{(\alpha)} $, where $U_{pq}^{(\alpha)}$ is an $N \times N$ matrix representation of $\hat{U}_{\alpha}$ \cite{Kivlichan_Babbush:2018, Yen_Izmaylov:2021, Cohn_Parrish:2021}. Measuring $\hat{H}_{\alpha}^{\prime}$ instead of $\hat{H}_{\alpha}$ gives the same $E$ because repartitioning does not change the operator sum: $\sum_{\alpha=0}^{N_{f}}\hat{H}_{\alpha}^{\prime} = \sum_{\alpha=0}^{N_{f}}\hat{H}_{\alpha} = \hat{H}$. 
In contrast, $M(\epsilon)$ changes with the choice of $c_{p}^{(\alpha)}$ because $\mathrm{Var}_{\psi}(\hat{H}_{\alpha}^{\prime})$ has a non-linear dependence on $c_{p}^{(\alpha)}$. 
As a consequence, one can reduce $M(\epsilon)$ by optimizing $c_{p}^{(\alpha)}$. 
Linearity of fermionic fragments with respect to $c_{p}^{(\alpha)}$ makes 
variance optimization particularly efficient.

\subsection{Optimization of the number of measurements}
\label{subsec:optimal_repartitioning}

In the following, we will present the approach for optimally repartitioning the one- and two-electron fragments to lower $M(\epsilon)$ (initial fragments are obtained as described in \hyperref[appendixa]{Appendix~A}). 
Since $M(\epsilon)$ depends on the fragment variances evaluated with the quantum wavefunction $|\psi\rangle$, which is classically difficult, we minimize the approximation to $M(\epsilon)$ computed with $\mathrm{Var}_{\phi}(\hat{H}_{\alpha}^{\prime})$: 
\begin{equation}
M_{\phi}(\epsilon) = \frac{1}{\epsilon^{2}} \sum_{\alpha=0}^{N_{f}} \frac{\mathrm{Var}_{\phi}(\hat{H}_{\alpha}^{\prime})}{m_{\alpha}};\label{eq:approx_M}
\end{equation}
in this work, the configuration interaction singles and doubles (CISD) wavefunction was used as the classically efficient proxy for the quantum wavefunction ($|\phi\rangle$).
The variances of the fragments after the repartition [$\hat{H}_{\alpha}^{\prime}$ in Eqs.~(\ref{eq:mod_oe}) and (\ref{eq:mod_te})] are obtained as 
\begin{align}
    \mathrm{Var}_{\phi}(\hat{H}_{0}^{\prime}) = &\mathrm{Var}_{\phi}(\hat{H}_{0}) \nonumber \\ &+ \sum_{\alpha,\beta}^{N_{f}}\sum_{p,q}^{N} c_{p}^{(\alpha)}c_{q}^{(\beta)} \mathrm{Cov}_{\phi}(\hat{O}_{p}^{(\alpha)}, \hat{O}_{q}^{(\beta)}) \nonumber \\ &+ \sum_{\alpha}^{N_{f}}\sum_{p}^{N} c_{p}^{(\alpha)} \overline{\mathrm{Cov}}_{\phi}(\hat{H}_{0}, \hat{O}_{p}^{(\alpha)}), 
\end{align}
\begin{align}
    \mathrm{Var}_{\phi}(\hat{H}_{\alpha}^{\prime}) = &\mathrm{Var}_{\phi}(\hat{H}_{\alpha}) \nonumber \\ &+ \sum_{p,q}^{N} c_{p}^{(\alpha)}c_{q}^{(\alpha)} \mathrm{Cov}_{\phi}(\hat{O}_{p}^{(\alpha)}, \hat{O}_{q}^{(\alpha)}) \nonumber \\ &- \sum_{p}^{N} c_{p}^{(\alpha)} \overline{\mathrm{Cov}}_{\phi}(\hat{H}_{\alpha}, \hat{O}_{p}^{(\alpha)}),
\end{align}
where we introduced $\overline{\mathrm{Cov}}_{\phi}(\hat{A},\hat{B}) = \mathrm{Cov}_{\phi}(\hat{A},\hat{B}) + \mathrm{Cov}_{\phi}(\hat{B},\hat{A})$ and $\hat{O}_{p}^{(\alpha)} = \hat{U}_{\alpha}^{\dagger} \hat{n}_{p} \hat{U}_{\alpha}$ for notational simplicity. To minimize Eq.~(\ref{eq:approx_M}) with respect to $c_{p}^{(\alpha)}$ and $m_{\alpha}$, we perform two-step iterative optimization following Refs.~\citenum{Yen_Izmaylov:2022} and \citenum{Choi_Izmaylov:2022}: 1) $c_{p}^{(\alpha)}$ are optimized with fixed $m_{\alpha}$ and 2) $m_{\alpha}$ are updated to the optimal allocation according to Eq.~(\ref{eq:m_alpha_opt}) with fixed $c_{p}^{(\alpha)}$. As this iterative procedure is a linearization heuristic for the non-linear problem of simultaneously optimizing $c_{p}^{(\alpha)}$ and $m_{\alpha}$, the obtained solution does not necessarily correspond to the minimum. Yet, in practice, we could typically reach convergence within $\sim 5$ iterations to find solutions with low gradients. To optimize the $c_{p}^{(\alpha)}$ variables with fixed $m_{\alpha}$, one can solve the system of linear equations:
\begin{align}
\epsilon^{2}\frac{\partial M_{\phi}(\epsilon)}{\partial c_{p}^{(\alpha)}} &= \frac{1}{m_{0}} \Big[ \sum_{\beta}^{N_{f}} \sum_{q}^{N}c_{q}^{(\beta)} \overline{\mathrm{Cov}}_{\phi}(\hat{O}_{p}^{(\alpha)}, \hat{O}_{q}^{(\beta)}) \nonumber \\ & \hspace{75pt} + \overline{\mathrm{Cov}}_{\phi}(\hat{H}_{0},\hat{O}_{p}^{(\alpha)}) \Big] \nonumber \\ &\hspace{15pt} + \frac{1}{m_{\alpha}} \Big[ \sum_{q}^{N} c_{q}^{(\alpha)} \overline{\mathrm{Cov}}_{\phi}(\hat{O}_{p}^{(\alpha)},\hat{O}_{q}^{(\alpha)})  \nonumber \\ &\hspace{38pt}- \overline{\mathrm{Cov}}_{\phi}(\hat{H}_{\alpha}, \hat{O}_{p}^{(\alpha)})\Big] = 0. \label{eq:lin_eq_full}
\end{align}
The final $m_{\alpha}$ obtained at the end of the iterative procedure suggests the optimal allocation of the total budget for each $\hat{H}_{\alpha}$. The suggested $M m_{\alpha}$ measurements for each fragment $\hat{H}_{\alpha}$ is not an integer, but rounding $M m_{\alpha}$ to the nearest integer should only have a negligible effect on the measurement error because $M \gtrsim 10^{6}$ in practice. 
We will refer to the algorithm proposed in this work as the fluid fermionic fragments (F$^3$) method.

\subsection{\label{subsec:reducing_c}Reducing the number of optimization variables}

The computational cost for optimizing $c_{q}^{(\alpha)}$ in the F$^3$ method increases with the number of $c_{q}^{(\alpha)}$ ($N_{c}$). The two main contributors to the computational time are the evaluation of covariances, $\overline{\mathrm{Cov}}_{\phi}(\hat{O}_{p}^{(\alpha)}, \hat{O}_{q}^{(\beta)})$, and solving the system of linear Eqs.~(\ref{eq:lin_eq_full}). Because the evaluation of $\overline{\mathrm{Cov}}_{\phi}(\hat{O}_{p}^{(\alpha)}, \hat{O}_{q}^{(\beta)})$ scales quadratically with $N_{c}$ and the computational time required for solving Eq.~(\ref{eq:lin_eq_full}) has an approximately cubic scaling with $N_{c}$, the cost of F$^3$ can be lowered significantly by reducing $N_{c}$. Therefore, we propose several restrictions on $c_{q}^{(\alpha)}$ to lower their number. 

Using spin symmetry of the electronic Hamiltonian written in a spin-restricted spin-orbital basis, 
we achieve a twofold reduction in the number of $c_{q}^{(\alpha)}$. Note that 
$\lambda_{2i-1,2i-1}^{(\alpha)} = \lambda_{2i,2i}^{(\alpha)}$ for $i = 1, \dots, N/2$ in the initial 
$\hat{H}_{\alpha}$ fragments obtained by considering the smaller $\tilde{g}_{ijkl}$ tensor over spatial orbitals 
(see \hyperref[appendixa]{Appendix~A} for definitions). Because $\lambda_{2i-1,2i-1}^{(\alpha)} = \lambda_{2i,2i}^{(\alpha)}$, we impose that the same amount of $\lambda_{2i-1,2i-1}^{(\alpha)}$ and $\lambda_{2i,2i}^{(\alpha)}$ is extracted from $\hat{H}_{\alpha}$, i.e., we impose that $c_{2i-1}^{(\alpha)} = c_{2i}^{(\alpha)} = \tilde{c}_{i}^{(\alpha)}$, thereby reducing the number of optimization variables by half. 
In the resulting F$^3$-Full method, $N_{c} = N_{f}N/2$, and the system of equations is simplified to
\begin{align}
\epsilon^{2}\frac{\partial M_{\phi}(\epsilon)}{\partial \tilde{c}_{i}^{(\alpha)}} &= \frac{1}{m_{0}} \Big[ \sum_{\beta}^{N_{f}} \sum_{j}^{N/2}\tilde{c}_{j}^{(\beta)} \overline{\mathrm{Cov}}_{\phi}(\hat{P}_{i}^{(\alpha)}, \hat{P}_{j}^{(\beta)}) \nonumber \\ & \hspace{75pt} + \overline{\mathrm{Cov}}_{\phi}(\hat{H}_{0},\hat{P}_{i}^{(\alpha)})\Big] \nonumber\\&\hspace{15pt} + \frac{1}{m_{\alpha}} \Big[ \sum_{j}^{N/2} \tilde{c}_{j}^{(\alpha)} \overline{\mathrm{Cov}}_{\phi}(\hat{P}_{i}^{(\alpha)},\hat{P}_{j}^{(\alpha)}) \nonumber \\ &\hspace{38pt} - \overline{\mathrm{Cov}}_{\phi}(\hat{H}_{\alpha}, \hat{P}_{i}^{(\alpha)})\Big] = 0, \label{eq:lin_v1}
\end{align}
where $\hat{P}_{i}^{(\alpha)} = \hat{O}_{2i-1}^{(\alpha)} + \hat{O}_{2i}^{(\alpha)} = \sum_{\sigma} \hat{U}_{\alpha}^{\dagger} \hat{n}_{i\sigma} \hat{U}_{\alpha}$.

Even more drastic reduction in $N_{c}$ can be achieved if we restrict $c_{p}^{(\alpha)}$ to be $p$-independent. This restricts us to repartitioning of only a fraction of the entire one-electron part of a two-electron $\hat{H}_{\alpha}$ fragment [i.e., a fraction of $\hat{U}_{\alpha}^{\dagger} ( \sum_{p} \lambda_{pp}^{(\alpha)} \hat{n}_{p})\hat{U}_{\alpha}$]. To this end, we restrict $c_{p}^{(\alpha)}$ as a scalar multiple of $\lambda_{pp}^{(\alpha)}$: $c_{p}^{(\alpha)} = c^{(\alpha)}\lambda_{pp}^{(\alpha)}$. As a result, $N_{c} = N_{f}$, and the system~(\ref{eq:lin_eq_full}) simplifies down to
\begin{align}
&\epsilon^{2}\frac{\partial M_{\phi}(\epsilon)}{\partial c^{(\alpha)}} = 
 \frac{1}{m_{0}} \Big[ \sum_{\beta}^{N_{f}}c^{(\beta)} \overline{\mathrm{Cov}}_{\phi}(\hat{O}^{(\alpha)}, \hat{O}^{(\beta)}) \nonumber \\ & \hspace{20pt} + \overline{\mathrm{Cov}}_{\phi}(\hat{H}_{0},\hat{O}^{(\alpha)})\Big]  + \frac{1}{m_{\alpha}} \Big[2 c^{(\alpha)} \mathrm{Var}_{\phi}(\hat{O}^{(\alpha)}) \nonumber \\ &  \hspace{85pt} - \overline{\mathrm{Cov}}_{\phi}(\hat{H}_{\alpha}, \hat{O}^{(\alpha)})\Big] = 0, \label{eq:lin_v2}
\end{align}
where $\hat{O}^{(\alpha)} = \sum_{p}^{N} \lambda_{pp}^{(\alpha)} \hat{O}_{p}^{(\alpha)}$. We will 
refer to this reduced version of F$^3$ as F$^3$-R1. 

Yet another reduction of variables can be done and is motivated by the relationship between 
$[\mathrm{Var}_{\psi}(\hat{H}_{\alpha})]^{1/2}$ appearing in the expression for the total measurement number with optimal $m_{\alpha}$ [$M_{\mathrm{opt}}(\epsilon)$ in \hyperref[appendixa]{Appendix~A}]  and the $L_{1}$ norm of a coefficient vector for a linear combination of unitaries (LCU) decomposition: $\hat{H}_{\alpha} = \sum_{j} a_{j}^{(\alpha)} \hat{V}_{j} + d_{\alpha} \hat{1}$, where $\hat{V}_{j}$ are some unitaries \cite{Loaiza_Izmaylov:2022}. Maximum $[\mathrm{Var}_{\psi}(\hat{H}_{\alpha})]^{1/2}$ for any $|\psi\rangle$ occurs when $|\psi\rangle = (|\mathrm{max}\rangle_{\alpha} + |\mathrm{min}\rangle_{\alpha})/\sqrt{2}$, where $|\mathrm{max}\rangle_{\alpha}$ ($|\mathrm{min}\rangle_{\alpha}$) is the eigenstate of $\hat{H}_{\alpha}$ with the highest (lowest) eigenvalue, $E_{\mathrm{max}}^{(\alpha)}$ ($E_{\mathrm{min}}^{(\alpha)}$); the corresponding maximum is 
\begin{equation}
\max_{\psi}\sqrt{\mathrm{Var}_{\psi}(\hat{H}_{\alpha})} = \Delta E_{\alpha}/2 \equiv (E_{\mathrm{max}}^{(\alpha)} - E_{\mathrm{min}}^{(\alpha)})/2.
\end{equation}
Using Theorem~1 of Ref.~\citenum{Loaiza_Izmaylov:2022}, which shows that the LCU $L_{1}$ norm, $\sum_{j} | a_{j}^{(\alpha)}|$, is in turn an upper bound for $\Delta E_{\alpha}/2$, we find that
\begin{equation}
\sqrt{\mathrm{Var}_{\psi}(\hat{H}_{\alpha})} \leq \sum_{j} | a_{j}^{(\alpha)}|. \label{eq:frag_ub}
\end{equation}
Thus, one can use $\hat{H}_{\alpha}$ with a low LCU $L_{1}$ norm as a heuristic approach to lowering $M(\epsilon)$.

One way to reduce the $L_{1}$ norm for a collection of $\hat{H}_{\alpha}$ while maintaining their measurability is substituting every $\hat{n}_{p}$ operator in the two-electron $\hat{H}_{\alpha}$ fragments with reflections: $\hat{r}_{p} = 1-2\hat{n}_{p}$ (satisfying $\hat{r}_{p}^{2} = 1$ and $\hat{r}_{p}^{\dagger} = \hat{r}_{p}$) \cite{Berry_Babbush:2019, vonBurg_Troyer:2021, Lee_Babbush:2021, Loaiza_Izmaylov:2022}. Because $\hat{r}_{p}$ maps onto an all-$\hat{z}$ Pauli product under all standard transformations (e.g., it maps to $\hat{z}_{p}$ under the Jordan--Wigner transformation) \cite{Seeley_Love:2012}, the fragment remains measurable even after the substitution:
\begin{equation}
\hat{H}_{\alpha}^{\prime} = \hat{U}_{\alpha}^{\dagger}\left( \sum_{pq}^{N} \frac{\lambda_{pq}^{(\alpha)}}{4} \hat{r}_{p}\hat{r}_{q} \right)\hat{U}_{\alpha}.
\end{equation}
To ensure that $\sum_{\alpha=0}^{N_{f}} \hat{H}_{\alpha}^{\prime} = \hat{H}$, the one-electron $\hat{H}_{0}$ term must also be modified as
\begin{equation}
\hat{H}_{0}^{\prime} = \hat{H}_{0} + \sum_{\alpha=1}^{N_{f}} \sum_{pq}^{N} \left( \lambda_{pq}^{(\alpha)} \hat{O}_{p}^{(\alpha)} - \frac{\lambda_{pq}^{(\alpha)}}{4} \right). \label{eq:H_0_rflxn}
\end{equation}
Concerning measurements, the constant term in Eq.~(\ref{eq:H_0_rflxn}) affects neither $M(\epsilon)$ nor the measurability of the fragments. Moving the constant terms, $- \sum_{pq} \lambda_{pq}^{(\alpha)}/4$, from $\hat{H}_{0}^{\prime}$ back to each $\hat{H}_{\alpha}^{\prime}$ reveals the connection to F$^3$: the substitution of $\hat{n}_{p}$ with $\hat{r}_{p}$ simply corresponds to the F$^3$ approach with $c_{p}^{(\alpha)}$ fixed as $c_{p}^{(\alpha)}= \sum_{q} \lambda_{pq}^{(\alpha)}$. 

Inspired by this connection to F$^3$ and the reduction in the upper bound for the fragment variances achieved by the substitution of $\hat{r}_{p}$, we heuristically propose F$^3$-R2 that restricts the optimization variables as $c_{p}^{(\alpha)} = c^{(\alpha)} \sum_{q} \lambda_{pq}^{(\alpha)}$. Note that this choice is equivalent to substituting $\hat{r}_{p}$ only for a fraction of the two-electron fragment ($c^{(\alpha)}\hat{H}_{\alpha}$) while leaving the rest [$(1-c^{(\alpha)})\hat{H}_{\alpha}$] as functions of $\hat{n}_{p}$. The $c^{(\alpha)}$ variables in F$^3$-R2 are optimized by solving the system of equations identical to Eq.~(\ref{eq:lin_v2}) except that $\hat{O}^{(\alpha)} = \sum_{pq}^{N} \lambda_{pq}^{(\alpha)} \hat{O}_{p}^{(\alpha)}$ (instead of $\hat{O}^{(\alpha)} = \sum_{p}^{N} \lambda_{pp}^{(\alpha)} \hat{O}_{p}^{(\alpha)}$ in F$^3$-R1).

\section{Results and discussions}
\label{sec:results}

\begin{table*}[]
\centering
\caption{Required measurement numbers [$\epsilon^2 M(\epsilon)$] in the different versions of F$^3$ applied to either LR or FR fragments are compared with $\epsilon^2 M(\epsilon)$ in LR, GFRO, ICS \cite{Yen_Izmaylov:2022}, and SPP \cite{Choi_Izmaylov:2022} for Hamiltonians of several molecules ($N$ is the number of spin-orbitals and is equal to the number of qubits).
 \label{tab:I}}
\begin{tabular}{lccccccccccc}
\hline
\hline
 & & & & & & \multicolumn{3}{c}{F$^3$-LR}&\multicolumn{3}{c}{F$^3$-FR} \\
Sys & $N$ & LR & GFRO & ICS & SPP &  Full  & R1 & R2 & Full & R1& R2 \\ 
\hline
H$_{3}^{+}$& 6   &  0.458 & 0.261 & 0.179 & 0.142 & 0.148 & 0.156 & 0.162 & 0.0803 & 0.0815 & 0.0805 \\
H$_{4}$    & 8   &  1.50 & 0.643 & 0.754 & 0.446 & 0.538 & 0.578 & 0.554 & 0.316 & 0.318 & 0.317 \\
H$_{6}$    & 12  &  3.06 & 1.21 & 2.27 & 1.23 & 1.08 & 1.42 & 1.13 & 0.519 & 0.587 & 0.554 \\
HF         & 12  &  43.5 & 41.9 & 0.670 & 0.436 & 0.278 & 7.10 & 0.454 & 0.327 & 15.5 & 0.593 \\
LiH        & 12    &  3.16 & 2.73 & 0.295 & 0.148 & 0.127 & 0.338 & 0.196 & 0.122 & 0.197 & 0.165 \\
CH$_{2}$   & 14  &  22.0 & 15.6 & 2.75 & 0.796 & 0.985 & 4.80 & 1.23 & 0.689 & 12.2 & 0.830 \\
BeH$_{2}$  & 14  &  1.86 & 1.61 & 0.543 & 0.341 & 0.543 & 0.848 & 0.680 & 0.430 & 0.948 & 0.583 \\
H$_{2}$O   & 14  &  58.5 & 49.4 & 2.05 & 1.16 & 0.892 & 9.83 & 1.10 & 0.709 & 25.7 & 0.911 \\
NH$_{3}$   & 16  &  58.1 & 46.1 & 4.83 & 2.62 & 1.49 & 9.22 & 1.70 & 0.990 & 31.3 & 1.18 \\
\hline
\hline
\end{tabular}
\end{table*}

\begin{table}[]
\caption{Number of optimization variables ($N_{c}$) in the different versions of F$^3$ applied to either LR or FR fragments is compared with $N_{c}$ in ICS \cite{Yen_Izmaylov:2022} and SPP \cite{Choi_Izmaylov:2022} for the systems presented in Table~\ref{tab:I}. 
 \label{tab:II}}
\resizebox{\columnwidth}{!}{
\begin{tabular}{lccccccc}
\hline
\hline
 & & & & \multicolumn{2}{c}{F$^3$-LR}&\multicolumn{2}{c}{F$^3$-FR} \\
Sys & $N$ & ICS & SPP & Full  & R1/R2 & Full & R1/R2  \\ 
\hline
H$_{3}^{+}$& 6   &  93 & 97 & 18 & 6 & 24 & 8 \\
H$_{4}$    & 8   &  236 & 644 & 40 & 10 & 92 & 23 \\
H$_{6}$    & 12  &  1702 & 8002 & 108 & 18 & 354 & 59 \\
HF         & 12  &  1220 & 3409 & 126 & 21 & 384 & 64 \\
LiH        & 12    & 1346 & 2718 & 126 & 21 & 354 & 59 \\
CH$_{2}$   & 14  &   1765 & 7534 & 196 & 28 & 539 & 77 \\
BeH$_{2}$  & 14  &    1265 & 3048 & 196 & 28 & 602 & 86 \\
H$_{2}$O   & 14  &   1863 & 7987 & 196 & 28 & 938 & 134 \\
NH$_{3}$   & 16  &   7203 & 35875 & 288 & 36 & 904 & 113 \\
\hline
\hline
\end{tabular}
}
\end{table}
We compare $M(\epsilon)$ in the different versions of F$^3$ applied to either LR or FR fragments with $M(\epsilon)$ in the initial LR and GFRO fragments (see \hyperref[appendixa]{Appendix~A} for definitions). 
In addition, the performance of F$^3$ is compared with the best qubit-space techniques that have the lowest $M(\epsilon)$: the iterative coefficient splitting (ICS) \cite{Yen_Izmaylov:2022} and shared Pauli products (SPP) \cite{Choi_Izmaylov:2022} methods. In particular, it was shown in Ref.~\citenum{Yen_Izmaylov:2022} that for measuring the expectation values of molecular electronic Hamiltonians, the $M(\epsilon)$ values in ICS and SPP are severalfold lower than $M(\epsilon)$ in some recently developed classical-shadow-based techniques \cite{Huang_Preskill:2021, Wu_Yuan:2021}.

The algorithms were used to compute $\epsilon^2 M(\epsilon)$ for electronic Hamiltonians of several molecules in the STO-3G basis and the following nuclear geometries: $R(\mathrm{H} - \mathrm{H}) = 1$\AA\ with $\angle\mathrm{H}\mathrm{H}\mathrm{H}=180\degree$ (for H$_{3}^{+}$, H$_{4}$, and H$_{6}$), $R(\mathrm{H} - \mathrm{X}) = 1$\AA\ (for X$=$F, Li), $R(\mathrm{C} - \mathrm{H}) = 1$\AA\ with $\angle\mathrm{H}\mathrm{C}\mathrm{H}=101.9\degree$ (for CH$_{2}$), $R(\mathrm{Be} - \mathrm{H}) = 1$\AA\ with $\angle\mathrm{H}\mathrm{Be}\mathrm{H}=180\degree$ (for BeH$_{2}$), $R(\mathrm{O} - \mathrm{H}) = 1$\AA\ with $\angle\mathrm{H}\mathrm{O}\mathrm{H}=107.6\degree$ (for H$_{2}$O), and $R(\mathrm{N} - \mathrm{H}) = 1$\AA\ with $\angle\mathrm{H}\mathrm{N}\mathrm{H}=107\degree$ (for NH$_{3}$). The data for the electronic Hamiltonians [Eq.~(\ref{eq:Ham_spin_orb})] and the corresponding initial LR and GFRO fragments [Eq.~(\ref{eq:frag_ham})] can be found in Ref.~\citenum{Choi_Izmaylov_data:2022}.

The presented $\epsilon^2 M(\epsilon)$ value is equivalent to the number of measurements in millions required to obtain $E$ with $10^{-3}$ a.u. accuracy (see \hyperref[appendixb]{Appendix~B}). While the optimization in F$^3$ is performed using the covariances approximated with the CISD wavefunction, $|\phi\rangle$, once the optimal fragments and measurement allocation are obtained, the $\epsilon^2 M(\epsilon)$ values are then evaluated according to Eq.~(\ref{eq:M_req}) using the exact covariances computed with the quantum wavefunction, $|\psi \rangle$. [Note that the use of approximate covariances instead of the exact covariances in F$^3$ had an insignificant influence on $\epsilon^2 M(\epsilon)$. In all systems, $\epsilon^2 M(\epsilon)$ in F$^3$ optimized using the exact covariances were lower only by $9\%$ or less compared to the presented $\epsilon^2 M(\epsilon)$ values in F$^3$ optimized with approximate covariances.]

Table~\ref{tab:I} shows that between the different versions of F$^3$, the most flexible full version yields the lowest $M(\epsilon)$ in all examples: on average, $M(\epsilon)$ in F$^3$-Full is a factor of $11$ lower than that in F$^3$-R1 and a factor of $1.2$ lower than that in F$^3$-R2. However, as shown in Table~\ref{tab:II}, the increased flexibility also results in F$^3$-Full having $N/2$-fold more optimization variables than the other versions. While F$^3$-R1 and F$^3$-R2 have the same $N_{c}$, F$^3$-R2 has a much lower $M(\epsilon)$ for many of the molecules; for the others, it has a similar $M(\epsilon)$ to that in F$^3$-R1. The success of F$^3$-R2 can be heuristically justified since it is designed to lower the LCU $L_{1}$ norm, which is an upper bound for fragment variances, as discussed in Sec.~\ref{subsec:reducing_c}. Because F$^3$-R2 can achieve a lower $M(\epsilon)$ with the identical computational cost as F$^3$-R1, there is no reason to employ F$^3$-R1 instead of F$^3$-R2. Therefore, we omit F$^3$-R1 in the following discussion.

The comparison of $M(\epsilon)$ in F$^3$ with that in the initial set of 
LR or FR fragments demonstrates the success of the proposed method. On average, $M(\epsilon)$ in F$^3$-Full is a factor of $35$ lower than that in the initial fragments, 
and $M(\epsilon)$ in F$^3$-R2 is a factor of $24$ lower than that in the initial fragments. Since
$M(\epsilon)$ in GFRO is always lower than that in LR, $M(\epsilon)$ in F$^3$ is also typically lower when FR fragments are used as the initial fragments. However, finding the initial GFRO fragments involves an iterative non-linear optimization procedure which is computationally more expensive than the LR decomposition. Moreover, while $N_{f}$ in LR is upper bounded by ($N^{2}/8 + N/4$), $N_{f}$ in GFRO is usually higher (for the presented molecules, GFRO has, on average, three times more fragments). 
Since the number of optimization variables is directly proportional to 
$N_{f}$ ($N_{c} = N_{f}N/2$ in F$^3$-Full and $N_{c} = N_{f}$ in F$^3$-R2), 
employing F$^3$ on the FR fragments requires more computational effort 
than employing it on the LR fragments.

For systems in Table~\ref{tab:II}, $N_{c}$ scales as $N^{x}$, where $x=2.8$ (for F$^3$-LR-Full), $x = 1.8$ (for F$^3$-LR-R2), $x = 3.7$ (for F$^3$-FR-Full), and $x = 2.7$ (for F$^3$-FR-R2). Because the bottleneck in the F$^3$ method is solving the system of linear equations [i.e., solving Eq.~(\ref{eq:lin_v1}) or (\ref{eq:lin_v2})], which has $\sim N_{c}^{3}$ scaling, the classical optimization costs in the different versions of F$^{3}$ have approximately $N^{8.5}, N^{5.5}, N^{11},$ and $N^{8.2}$ scaling. Therefore, F$^3$-LR-R2, with its classical computational cost scaling as $\sim N^{5.5}$ with system size ($N$), would be particularly useful for larger systems. 

Comparison of F$^3$ with state-of-the-art qubit-space techniques (ICS and SPP) shows that for all molecules except BeH$_{2}$, F$^3$ has lower $M(\epsilon)$. In particular, even the computationally most efficient combination of applying F$^3$-R2 to the LR fragments yields a lower $M(\epsilon)$ than that in ICS, whereas this most efficient combination yields a similar $M(\epsilon)$ to that in SPP. Furthermore, $M(\epsilon)$ in the best fermionic-space technique (F$^3$-FR-Full) is, on average, a factor of 1.6 smaller than $M(\epsilon)$ in the best qubit-space technique (SPP). The bottleneck in the optimization procedures in all three methods (F$^3$, ICS, and SPP) is solving a system of linear equations. Therefore, one can assess the required computational cost by examining the number of optimization variables. Table~\ref{tab:II} shows that even the most computationally costly combination, F$^3$-FR-Full 
has a much lower $N_{c}$ than that in either ICS or SPP. The classical computational costs of ICS and SPP have approximately $N^{12}$ and $N^{15}$ scaling, which are worse even than that in F$^3$-FR-Full, with the worst scaling among different versions of F$^3$.

\section{Conclusion}
\label{sec:conclusion}
This work proposes a new method that achieves a significant reduction in the number of measurements by taking advantage of the possibility of extracting fractions of one-electron parts from the two-electron fermionic fragments and combining them with the purely one-electron fragment. This repartitioning keeps  the fragments measurable and conserves the Hamiltonian expectation value. 
On the other hand, the number of measurements due to its dependence on variances of fragments 
can be lowered by this repartitioning. The proposed algorithm finds the repartitioning that minimizes 
the number of measurements and achieves a severalfold reduction in the number of measurements
compared to those for initially generated fragments.

Even though the number of measurements in the previously proposed methods suggested that the qubit-space techniques are superior to their fermionic counterparts, by employing the fluid fermionic fragments  method, we were able to achieve the number of measurements lower even than those in the best qubit-space techniques (ICS \cite{Yen_Izmaylov:2022} and SPP \cite{Choi_Izmaylov:2022}). 
Furthermore, compared to these techniques, the method presented here was shown to have much fewer optimization variables. In particular, the number of optimization variables in the computationally 
most efficient version of the proposed algorithm scales sub-quadratically with the number of 
spatial orbitals, thereby making the algorithm applicable for larger systems. In addition, while the errors due to the non-unit quantum gate fidelities were not considered in this work, fermionic fragments 
conserve molecular symmetries and their measurements can be corrected by error mitigation methods
employing these symmetries \cite{Bonet-Monroig_OBrien:2018, Endo_Yuan:2021, Cai_OBrien:2022}. Therefore, the advantage of the proposed fluid fermionic fragments method over the qubit-space techniques should be even greater when quantum hardware errors are taken into account.

While this work focused on the measurements in VQE for ground-state energy estimation, some promising extensions of VQE, including those for excited-state calculations \cite{McClean_deJong:2017, Huggins_Whaley:2020, Stair_Evangelista:2020}, require measuring expectation values of additional effective Hamiltonians containing three- and four-electron terms. A common approach for lowering the number of measurements required to obtain these additional expectation values relies on improving the efficiency of simultaneously estimating all three- and four-body reduced density matrices \cite{Bonet-Monroig_OBrien:2020, Zhao_Miyake:2021}. Alternatively, as a potential extension of this work, one could develop a more targeted algorithm that finds optimal fragments for only a few necessary effective Hamiltonians. It would be informative to examine whether this more targeted approach can further reduce the required number of measurements.

Lastly, although we employed the fluid fermionic fragments method to optimize quantum measurements, it can also be used to optimize Hamiltonian fragments for other purposes. In the fault-tolerant paradigm, one can solve the electronic structure problem by time evolving the initial guess state (with a good overlap with the ground state of $\hat{H}$), then applying the quantum phase estimation algorithm \cite{Aspuru-Guzik_Head-Gordon:2005}. A commonly employed technique for the time evolution is Trotterization \cite{Lloyd:1996, Martinez-Martinez_Izmaylov:2022}. Because the Hamiltonian fragments required for Trotterization are equivalent to the measurable fragments considered in this work, the fluid fermionic fragments method can also be applied in the context of Trotterization. The only necessary modification is that instead of minimizing the number of measurements required in a VQE step, one would employ the fluid fermionic fragments method to minimize the Trotter error \cite{Martinez-Martinez_Izmaylov:2022, Suzuki:1990}.

\section*{Acknowledgments}
A.F.I. is grateful to Tom O'Brien for insightful discussions and acknowledges financial support from the Google Quantum Research Program, Zapata Computing Inc.,  and the National Science Foundation under Grant No. NSF PHY-1748958. S.C. thanks Tzu-Ching Yen for helpful discussions and acknowledges financial support from the Swiss National Science Foundation through the Postdoc Mobility Fellowship (Grant No. P500PN-206649). This research was enabled in part by support provided by Compute Ontario and Compute Canada.

\section*{Appendix A: Initial Hamiltonian fragments}
\label{appendixa}
 
 The LR decomposition is less ambiguous compared to its FR counterpart, 
 and we use an LR decomposition procedure described in Ref.~\citenum{Huggins_Babbush:2021}. 
  Among different implementations of the FR decomposition, 
 we employ the ``greedy'' FR optimization (GFRO), 
since it has the lowest $M(\epsilon)$ \cite{Yen_Izmaylov:2021}.
Greedy algorithms typically have low $M(\epsilon)$, and their success can be attributed to the sum of square roots appearing in
\begin{equation}
M_{\rm opt}(\epsilon) = \frac{1}{\epsilon^{2}} \left[\sum_{\alpha=0}^{N_{f}} \sqrt{\mathrm{Var}_{\psi}(\hat{H}_{\alpha})}\right]^{2}\label{eq:M_opt}
\end{equation}
obtained by choosing the optimal measurement allocation,
\begin{equation}
 m_{\alpha} = \frac{\sqrt{\mathrm{Var}_{\psi}(\hat{H}_{\alpha})}}{\sum_{\beta=0}^{N_{f}}\sqrt{\mathrm{Var}_{\psi}(\hat{H}_{\beta})}},\label{eq:m_alpha_opt}
\end{equation}
that minimizes $M(\epsilon)$. For a fixed sum of $\mathrm{Var}_{\psi}(\hat{H}_{\alpha})$, the sum of square roots in Eq.~(\ref{eq:M_opt}) is lower if the variances are distributed unevenly, and greedy approaches tend to yield an uneven distribution of $\mathrm{Var}_{\psi}(\hat{H}_{\alpha})$. 

The computational effort of both LR and FR decompositions is reduced significantly by working with a smaller two-electron tensor over spatial orbitals, $\tilde{g}_{ijkl}$, instead of the tensor over spin-orbitals, $g_{pqrs}$. One can rewrite the electronic Hamiltonian as
\begin{align}
\hat{H} &= \sum_{\sigma}\sum_{ij}^{N/2} \tilde{h}_{ij} \hat{E}_{i\sigma}^{j\sigma} + \sum_{\sigma \sigma^{\prime}} \sum_{ijkl}^{N/2} \tilde{g}_{ijkl} \hat{E}_{i\sigma}^{j\sigma} \hat{E}_{k\sigma^{\prime}}^{l\sigma^{\prime}} \label{eq:Ham_orb} 
\end{align}
where $\sigma$ and $\sigma^{\prime}$ specify the spin-$z$ projection, 
while $\tilde{h}_{ij}$ and $\tilde{g}_{ijkl}$ are one- and two-electronic integrals:
\begin{align}
\tilde{h}_{ij} = \int d\vec{r} \phi_{i}^{\ast}(\vec{r}) \left(- \frac{\nabla^{2}}{2} - \sum_{I} \frac{Z_{I}}{| \vec{r} - \vec{r}_{I} |} \right) & \phi_{j}(\vec{r}) \nonumber \\& \hspace{-40pt} - \sum_{k}^{N/2} \tilde{g}_{ikkj}
\end{align}
and
\begin{equation}
\tilde{g}_{ijkl} = \frac{1}{2} \int \int d\vec{r}_{1}d\vec{r}_{2} \frac{\phi_{i}^{\ast}(\vec{r}_{1}) \phi_{j}(\vec{r}_{1}) \phi_{k}(\vec{r}_{2}) \phi_{l}^{\ast}(\vec{r}_{2})}{|\vec{r}_{1} - \vec{r}_{2}|},
\end{equation}
where $\phi_{i}(\vec{r})$ is the $i$th one-particle electronic basis function in the position representation, and the charge and position of the $I$th nucleus are denoted by $Z_{I}$ and $\vec{r}_{I}$. 

To optimize the computational cost for LR and FR decompositions, we work with $\tilde{g}_{ijkl}$ and 
then subsequently convert the resulting $\tilde{\lambda}_{ij}^{(\alpha)}$ and $\tilde{U}_{ij}^{(\alpha)}$ into $\lambda_{pq}^{(\alpha)}$ and ${U}_{pq}^{(\alpha)}$ according to
\begin{equation}
\lambda_{2i-1,2j-1}^{(\alpha)} = \lambda_{2i-1,2j}^{(\alpha)} = \lambda_{2i,2j-1}^{(\alpha)} = \lambda_{2i,2j}^{(\alpha)} = \tilde{\lambda}_{ij}^{(\alpha)},  
\end{equation}
\begin{align}
U_{2i-1,2j-1}^{(\alpha)} &= U_{2i,2j}^{(\alpha)} = \tilde{U}_{ij}^{(\alpha)}, \nonumber \\  U_{2i,2j-1}^{(\alpha)} &= U_{2i-1,2j}^{(\alpha)} = 0
\end{align}
for $i,j = 1, \dots, N/2$. 

The LR decomposition is particularly efficient because it finds the rank-1 matrices $\tilde{\lambda}_{ij}^{(\alpha)} = \tilde{\eta}_{i}^{(\alpha)}\tilde{\eta}_{j}^{(\alpha)}$ by diagonalizing the two-electron tensor $\tilde{g}_{ij,kl}$ considered as a matrix with each dimension spanned by a pair of basis indices. This diagonalization gives a theoretical limit on $N_{f} \leq N_{o}(N_{o}+1)/2$, where $N_{o} = N/2$, and the less-than sign originates from a truncation of the expansion by removing terms for low magnitude eigenvalues (see Ref.~\citenum{Motta_Chan:2021} for further details on LR decomposition). Having a low $N_{f}$ is beneficial for the fluid fermionic fragments method because the number of optimization variables ($c_{p}^{(\alpha)}$) is directly proportional to $N_{f}$. 

The classical computational cost of LR has $\sim N_{o}^{6}$ scaling due to the cost of diagonalization, whereas the precise scaling for GFRO cannot be determined since it is a heuristic algorithm. In the examples considered in this work, the classical cost of GFRO was comparable to that of the F$^3$ optimization. Typically, the LR decomposition requires less computational effort than GFRO, but $M(\epsilon)$ in GFRO is lower owing to more flexible $\tilde{\lambda}_{ij}^{(\alpha)}$ \cite{Yen_Izmaylov:2021, Cohn_Parrish:2021}. 

The $\alpha$th GFRO fragment, $\hat{H}_{\alpha}$, is found by minimizing the $L_{1}$ norm of the $\tilde{\mathbf{G}}^{(\alpha+1)}$ tensor in
\begin{equation}
\sum_{ijkl\sigma\sigma^{\prime}} \tilde{G}_{ijkl}^{(\alpha)} \hat{E}_{i\sigma}^{j\sigma} \hat{E}_{k\sigma^{\prime}}^{l\sigma^{\prime}} - \hat{H}_{\alpha} = \sum_{ijkl\sigma\sigma^{\prime}} \tilde{G}_{ijkl}^{(\alpha+1)} \hat{E}_{i\sigma}^{j\sigma} \hat{E}_{k\sigma^{\prime}}^{l\sigma^{\prime}},
\end{equation}
where $\tilde{G}_{ijkl}^{(1)} = \tilde{g}_{ijkl}$. In each iteration, the $L_{1}$ norm of $\tilde{\mathbf{G}}^{(\alpha+1)}$ is minimized over the space of $\{\tilde{\lambda}_{ij}^{(\alpha)}, \tilde{\theta}^{(\alpha)}\}$ variables parameterizing the $\alpha$th fragment,
\begin{equation}
\hat{H}_{\alpha} = \hat{U}(\tilde{\theta}^{(\alpha)})^{\dagger} \left( \sum_{ij\sigma\sigma^{\prime}} \tilde{\lambda}_{ij}^{(\alpha)} \hat{n}_{i\sigma} \hat{n}_{j\sigma^{\prime}} \right) \hat{U}(\tilde{\theta}^{(\alpha)}),
\end{equation}
where $\hat{U}(\tilde{\theta}^{(\alpha)}) = \exp[ \sum_{i > j}^{N_{o}} \tilde{\theta}_{ij}^{(\alpha)}\sum_{\sigma} (\hat{E}_{i\sigma}^{j\sigma} - \hat{E}_{j\sigma}^{i\sigma})]$. The iteration terminates when the $L_{1}$ norm of $\tilde{\mathbf{G}}^{\alpha+1}$ is below a given threshold ($1\cdot 10^{-5}$ is used in this work).

\section*{Appendix B: Number of measurements required to obtain the Hamiltonian expectation value}
\label{appendixb}
In every method considered in Sec.~\ref{sec:results}, the expectation value of the Hamiltonian, $E = \langle \psi | \hat{H} | \psi \rangle$, is obtained as the sum of expectation values of the separately measured $\hat{H}_{\alpha}$ fragments [see Eq.~(\ref{eq:measurable_frags})]. Therefore, the most straightforward estimator for $E$ (denoted by $\bar{H}$) is the sum of estimators for individual fragments, i.e., $\bar{H} = \sum_{\alpha} \bar{H}_{\alpha}.$
Each $\bar{H}_{\alpha}$ is the average result of $M_{\alpha}$ repeated quantum measurements of $\hat{H}_{\alpha}$:
\begin{equation}
\bar{H}_{\alpha} = \frac{1}{M_{\alpha}} \sum_{i=1}^{M_{\alpha}} H_{\alpha,i},
\end{equation}
where $H_{\alpha, i}$ is the $i$th outcome of measuring $\hat{H}_{\alpha}$.

Because each fragment is measured independently, the covariances between the fragments, $\mathrm{Cov}(\bar{H}_{\alpha}, \bar{H}_{\beta})$, are zero, and therefore the variance of $\bar{H}$ is simply the sum of variances of $\bar{H}_{\alpha}$:
\begin{equation}
\mathrm{Var}(\bar{H}) = \sum_{\alpha} \mathrm{Var}(\bar{H}_{\alpha}).
\end{equation}
Assuming that each $M_{\alpha}$ is large, we can invoke the central limit theorem to evaluate $\mathrm{Var}(\bar{H}_{\alpha})$ using the quantum operator variances, i.e., $\mathrm{Var}(\bar{H}_{\alpha}) = \mathrm{Var}_{\psi} (\hat{H}_{\alpha}) / M_{\alpha}$. Therefore, the Hamiltonian estimator variance can be evaluated as 
\begin{equation}
\mathrm{Var}(\bar{H}) = \sum_{\alpha} \frac{\mathrm{Var}_{\psi} (\hat{H}_{\alpha})}{M_{\alpha}} . \label{eq:estimator_variance}
\end{equation}
To obtain the expression for $M(\epsilon)$, i.e., the total number of required measurements to obtain $E$ with $\epsilon \equiv [\mathrm{Var}(\bar{H})]^{1/2}$ accuracy, we introduce fractional measurement allocation, $m_{\alpha} = M_{\alpha}/M$ (with $\sum_{\alpha} m_{\alpha} = 1$), then rearrange Eq.~(\ref{eq:estimator_variance}) as
\begin{equation}
M(\epsilon) = \frac{1}{\epsilon^{2}} \sum_{\alpha} \frac{\mathrm{Var}_{\psi}(\hat{H}_{\alpha})}{m_{\alpha}}.
\end{equation}

\bibliographystyle{quantum}
\bibliography{FFF}

\end{document}